\documentclass{article}
\usepackage{amsmath,amssymb,amsfonts,amsthm}

\newtheorem{theorem}{Theorem}[section]
\newtheorem{lemma}[theorem]{Lemma}
\newtheorem{definition}[theorem]{Definition}

\title{Initial data for stationary space-times near space-like infinity}
\author{Sergio Dain\\ 
Max-Planck-Institut f\"ur Gravitationsphysik\\
Am M\"uhlenberg 1\\
14476 Golm\\
Germany}

\begin{document}
\maketitle

\begin{abstract}
  We study  Cauchy initial data for asymptotically flat,
  stationary vacuum space-times near space-like infinity. The fall-off
  behavior of the intrinsic metric and the extrinsic curvature is
  characterized.  We prove that they have an analytic expansion in
  powers of a radial coordinate. The coefficients of the expansion are
  analytic functions of the angles.  This result allow us to fill a
  gap in the proof found in the literature of the statement that all
  asymptotically flat, vacuum stationary space-times admit an analytic
  compactification at null infinity.

  Stationary initial data are physical important and highly
  non-trivial examples of a large class of data with similar
  regularity properties at space-like infinity, namely, initial data
  for which the metric and the extrinsic curvature have asymptotic
  expansion in terms of powers of a radial coordinate. We isolate the
  property of the stationary data which is responsible for this kind
  of expansion.

\vspace{0.5cm}

\end{abstract}

\section{Introduction}
The far-field behavior of stationary space-times is by now reasonably
well understood (see \cite{Beig00b} and reference therein). However,
all of this information is encoded in terms of quantities intrinsic to
the abstract three dimensional manifold $\tilde X$ of trajectories of
the time like Killing vector $\xi^a$. In order to extract from this
description general properties of asymptotically flat solutions of the
field equations, we have to express this information in terms of
quantities that can be defined for general asymptotically flat
solutions. Take an space like hypersurface $\tilde S$ in the space
time.  Consider the intrinsic metric and the extrinsic curvature of
$\tilde S$. How are these fields related to the ones defined on the
abstract manifold $\tilde X$?  And, more important, what is the
precise fall off behavior of these fields at space-like infinity?  The
purpose of this article is to answer these questions.  The result is
somehow unexpected: in contrast the fields defined in $\tilde X$ the
fields in the Cauchy initial data $\tilde S$ are never analytic at
infinity, unless we are dealing with the static case. By `analytic at
infinity' we mean that the tensor components of the fields are
analytic with respect to an appropriate Cartesian coordinate system in
a neighborhood of infinity.  However, they are analytic in terms of a
radial coordinate and the corresponding angles.  The typical behavior
of the fields is represented by the radial coordinate
$r=(\sum_{i=1}^3(x^j)^2 )^{1/2}$ itself, which is not an analytic
function of the Cartesian coordinates $x^i$ at the origin. This is
precisely the non-analytic type of behavior of the intrinsic metric
and the extrinsic curvature near space like infinity in a Cauchy
slice.  The fall-off of those tensors in a particular foliation is
given by theorems \ref{tm} and \ref{tc}.  This constitute our main
result. These theorems 
 are essential in order to
prove that all stationary, asymptotically flat, space-times admit an
analytic conformal compactification of null infinity. We fill in this
way a gap in the proof of this result given in \cite{Damour90}.

In a previous work, we have studied  a class of initial data that have
asymptotic expansion in terms of powers of a radial coordinate\cite{Dain99}. In
order to construct this class we impose a condition on the
square of the conformal extrinsic curvature.   In theorem \ref{tc} we
prove that the stationary initial data satisfy the same condition.

\section{Initial data for stationary space-time near space like
  infinity}

Let $\tilde M, \tilde g_{ab}$ be a stationary vacuum space-time.  The
collection of all trajectories of the time like Killing vector $\xi^a$
defines an abstract manifold $\tilde X$, called `the quotient space'.
When $\xi^a$ is surface orthogonal (i.e. when the space-time is
static) then $\tilde X$ is naturally identified with one of the
hypersurfaces of $\tilde M$  everywere orthogonal to $\xi^a$.
Each trajectory of $\xi^a$ would intersect the hypersurface in exactly
one point. In the non-hypersurface orthogonal case, however, there is
no natural way of introducing such surface on $\tilde M$.  This is the
reason why $\tilde X$ and not a slice $\tilde S$ is the most
convenient object to study in the stationary
spaces-times \cite{Geroch71}.

The field equations have a
remarkably simple form in terms of quantities defined on $\tilde
X$. 
The  metric, with signature $+---$, can
locally be written as 
\begin{equation}
  \label{eq:stmetric}
  \tilde g = \lambda(dt+\beta_i d\tilde x^i)-\lambda^{-1} \tilde
  \gamma_{ij}d\tilde x^i
  d\tilde x^j \quad (i,j= 1,2,3),
\end{equation}
where $\lambda$, $\beta_i$, and the Riemannian metric $\tilde\gamma_{ij}$
depend only on the spatial coordinates $\tilde x^k$. In these
coordinates,  the Killing vector 
  is given by $\xi^a=(\partial/\partial t)^a$. Note that
$\lambda=g_{ab}\xi^a\xi^b$. 
The metric $\tilde\gamma_{ij}$,  $\lambda$ and  $\beta_i$ are
naturally defined on $\tilde X$.
 
On the other hand, the metric $\tilde g_{ab}$ has a 3+1
decomposition with respect to the hypersurface $\tilde S$,  defined
by $t=constant$,  
\begin{equation}
  \label{eq:ls}
  \tilde g = \tilde N^2 dt^2-\tilde h_{ij} (N^i dt+ d\tilde x^i )(N^j
  dt+ d\tilde x^j), 
\end{equation}
where $\tilde N$ is the lapse function,   $N^i$ the shift vector and $\tilde
h_{ij}$ the intrinsic metric of the slice $\tilde S$.
 We have the following relations between these quantities
\begin{equation}
  \label{eq:tN}
  \tilde N^2=\frac{\lambda}{1-\lambda^2 \tilde \beta^i \beta_i},
\end{equation}
\begin{equation}
  \label{eq:S}
  N_j=\lambda \beta_j, \quad N^j=-\tilde N^2 \lambda \tilde \beta^j,
\end{equation}
\begin{equation}
  \label{eq:th}
  \tilde h_{ij}=\lambda^{-1} \tilde \gamma_{ij} -\lambda \beta_i
  \beta_j, \quad \tilde h^{ij}=\lambda \tilde \gamma^{ij}
  +\lambda^2\tilde N^2 \beta^i
  \beta^j,
\end{equation}
were we have defined
\begin{equation}
  \label{eq:sb}
  \tilde N_j=\tilde h_{ij}N^i, \quad \tilde \beta^j=\tilde \gamma^{jk} \beta_k.
\end{equation}
Note that the indices are moved with different metrics. In general,
indexes of tensors defined on $\tilde X$ will be moved with the metric
$\tilde \gamma_{ij}$ and the ones defined on $\tilde S$ with $\tilde
h_{ij}$.

We will write now the vacuum field equations in terms of the quantities 
intrinsic to $\tilde X$. 
Let the  covariant derivative $\tilde 
D_i$ be  defined with respect to $\tilde
\gamma_{ij}$. From the vacuum field equations it follows that the quantity
\begin{equation}
  \label{eq:twist2}
  \omega_i = -\lambda^2 \tilde \epsilon_{ijk} \tilde D^j
  \beta^k ,
\end{equation}
 is curl free, i.e.
\begin{equation}
 \label{eq:cf}  
\tilde D_{[i} \omega_{j]}=0. 
\end{equation}
Where $\tilde \epsilon_{ijk}=\tilde \epsilon_{[ijk]}$ and $\tilde
\epsilon_{123}= |\det \tilde \gamma_{ij}|^{1/2}$. 
Thus, there exist a scalar field
  $\omega$ (the twist of the Killing vector $\xi^a$) such that
  \begin{equation}
    \label{eq:grad}
   \tilde D_i \omega=\omega_i.  
  \end{equation}
By equation \eqref{eq:twist2}, the covector $\beta_k$ is determined
only up to a gradient
\begin{equation}
  \label{eq:1}
  \beta_k \rightarrow \beta_k+\partial_k f,
\end{equation}
where $f$ is a scalar field in $\tilde X$. Under this change the
metric \eqref{eq:stmetric} remains unchanged if we set $t \rightarrow
t-f$.

It is convenient not to work with the scalar $\lambda$ and $\omega$
but with certain algebraic combinations  introduced
by Hansen \cite{Hansen74}
\begin{align}
  \label{eq:phiM}
  \tilde \phi_M &=\frac{1}{4\lambda}(\lambda^2 +\omega^2-1),\\
 \label{eq:phiS}
  \tilde \phi_S &=\frac{1}{2\lambda}\omega,\\
\label{eq:phiK}
  \tilde \phi_K &=\frac{1}{4\lambda}(\lambda^2 +\omega^2+1).
\end{align}
These functions  satisfy the relation
\begin{equation}
  \label{eq:rephi}
\tilde   \phi^2_M+\tilde   \phi^2_S-\tilde   \phi^2_K=-\frac{1}{4}. 
\end{equation}
Denote by $\tilde \phi$ any of the functions $\tilde \phi_M,\tilde
\phi_S,\tilde  \phi_K$.  The
vacuum field equations then read 
\begin{align}
  \label{eq:fieldphi}
  \tilde \Delta \tilde \phi &=2\tilde R \tilde \phi \\
\label{eq:fieldricci}
\tilde R_{ij} &=2 (\tilde D_i \tilde \phi_M \tilde D_j \tilde 
\phi_M+\tilde D_i \tilde \phi_S \tilde D_j \tilde \phi_S- \tilde D_i \tilde \phi_K \tilde D_j \tilde \phi_K),
\end{align}
where $\tilde R_{ij}$ is the Ricci tensor of $\tilde \gamma_{ij}$ and
$\Delta =\tilde D^i \tilde D_i$. Of
course, since (\ref{eq:rephi}) holds, only two of the three equations
(\ref{eq:fieldphi}) are independent.

We assume that   $(\tilde X, \tilde \gamma_{ij}, \tilde
\phi_M, \tilde \phi_S)$ is asymptotically flat in the
following sense. There exist a manifold $X$, consisting of $\tilde X$
and an additional point $i$; such that:

\begin{itemize}
\item[(i)]  For some real constant
$B^2>0$ the conformal factor
\begin{equation}
  \label{eq:cbs}
  \Omega=\frac{1}{2}B^{-2} [(1+4(\tilde \phi_M^2+\tilde\phi_S^2 ))^{1/2}-1  ]
\end{equation}
is $C^2$ on $X$ and satisfies
\begin{equation}
  \label{eq:Oi}
  \Omega(i)=0, \quad D_i \Omega(i)=0
\end{equation}
at the point $i$.

\item[(ii)] $\gamma_{ij}= \Omega^2 \tilde \gamma_{ij}$ extends to
  a $C^{4,\alpha}$ metric on $X$ and  
  \begin{equation}
    \label{eq:Oij}
    D_j D_k \Omega(i)= 2\gamma_{jk}(i).
  \end{equation}

\item[(iii)] $\Omega$ is  $C^{2,\alpha}$ on $X$. 

\end{itemize}

Define the rescaled functions  $\phi=\tilde \phi/\sqrt{\Omega}$. The
following theorem has been proved in \cite{Beig81}, see also \cite{Kundu81}.

\begin{theorem}[Beig-Simon] \label{bs}
For any asymptotically flat solution $(\tilde \gamma_{ij}, \tilde
\phi_M, \tilde \phi_S)$ of equations (\ref{eq:fieldphi}) and
(\ref{eq:fieldricci})  there exist a
chart defined in some neighborhood of $i$ in $X$ such that
$(\gamma_{ij}, \phi_M, \phi_S, \Omega)$ are analytic. 
\end{theorem}
The asymptotic flatness condition can be written in terms of the decay at
infinity of the  physical
metric $\tilde g_{ab}$. It can be relaxed  considerable, see
\cite{Kennefick95}. Existence of asymptotically solutions of the field 
equations, without any further symmetry,  has been proved in
\cite{Reula89}.   Theorem \ref{bs} make use of the  specific
conformal factor (\ref{eq:cbs}), other choices are also possible, for
example the one made in \cite{Kundu81}. 

Given the chart in which   $\gamma_{ij}$ is
analytic, we can make a   coordinate transformation to $\gamma$-normal
coordinates $x^j$ centered  at $i$. The fields  $\gamma_{ij}$,
$\phi_M$, $\phi_S$, $\Omega $
are also  analytic with respect to $x^j$.  We define the radial
coordinate  $r = (\sum_{i=1}^3 (x^j)^2 ) ^{1/2}$.  
The first terms in the expansion of $\phi_M$ and $\phi_S$ are given by
\begin{equation}
  \label{eq:sme}
  \phi_M=M+O(r), \quad \phi_S=S^ix_i + O(r^2),
\end{equation}
where  $M$ is the total mass, and $S^i$ is the intrinsic
angular momentum.

Is important to recall that theorem \ref{bs} asserts only the
analyticity of $\gamma_{ij}$, $\phi_M$, $\phi_S$, $\Omega $, and not
of the other quantities, like, for example, $\lambda, \phi_K,
\beta_i$. In fact these quantities are \emph{not} analytic as we will
see. However,
the non-analyticity of these functions  is given only by the function $r$. 
According to this, we define the following function space, which is
the analytic analog of the spaces defined in \cite{Dain99}.
\begin{definition}
We define the space $E^\omega$ as the set $E^\omega=\{f=f_1+rf_2 \, :
\, f_1,f_2 \text{ analytic functions in a neighborhood of } i \}$. 
\end{definition} 
We have the following lemma. 

\begin{lemma} \label{Eo}
Let $f, g \in E^\omega$, then 
\begin{itemize}
\item[(i)] $f+g \in E^\omega$

\item[(ii)] $fg\in E^\omega$

\item[(iii)] If $f\neq 0$ then $1/f \in E^\omega$
\end{itemize}
\end{lemma}
\begin{proof} The  first two assertions are obvious, for (iii) see
\cite{Dain99}. \end{proof}

In what follows, the idea is to prove that all the relevant fields
essentially belong to $E^\omega$.
We begin with  a useful lemma regarding the conformal
factor (\ref{eq:cbs}).
\begin{lemma} \label{r2}
The conformal factor $\Omega$ given by (\ref{eq:cbs}) has the following form
$\Omega=r^2 f_\Omega$, where $f_\Omega$ is an analytic,
positive, function, and $f_\Omega(i)=1$.
\end{lemma}
\begin{proof}
  The conformal factor $\Omega$  satisfies equation (16) of \cite{Beig81} (see
  also equation (2.9) of \cite{Kundu81}). This equation has the
  following form
\begin{equation}
  \label{eq:OBS}
  D_iD_j\Omega= \Omega T_{ij}+f_1\gamma_{ij}+f_2 D_i \Omega D_j\Omega,
\end{equation}
where the tensor $T_{ij}$ and the functions $f_1$, $f_2$ are analytic.
We want to prove that the symmetric and trace free part of the tensor
$D_{j_1} \cdots D_{j_n} \Omega$, for all $n$, vanish at the point $i$.
To prove this we use induction on $n$, in the same way as in the proof
of lemma 1 in \cite{Beig91b}. The cases   $n=1,2$ are given by equations
(\ref{eq:Oi}) and (\ref{eq:Oij}).  To perform the induction step we
assume $n\geq 2$ and show that the statement for $n-1$ implies that
for $n$. Using equation (\ref{eq:OBS}), we express  $D_{j_1} \cdots
D_{j_n} \Omega$ in terms of $D_{j_1} \cdots
D_{j_{n-1}} \Omega$. Using the  induction hypothesis the result
follows.  

We use the previous  result in the analytic expansion of  $\Omega$ 
\begin{equation}
  \label{eq:anO}
  \Omega=r^2+\frac{1}{3!}x^i x^j x^k D_i D_j D_k
  \Omega|_i+\frac{1}{4!}x^i x^j x^k x^qD_i D_j D_k D_q\Omega|_i \cdots,
\end{equation}
to conclude that $\Omega=r^2 f_\Omega$, where $f_\Omega$ is an
analytic, positive, function, and $f_\Omega(i)=1$.  
There exist a much more elegant method to prove the same
result directly out of equation (\ref{eq:OBS}) using complex
analysis, as it is explained in \cite{Friedrich88}. \end{proof}

From equation (\ref{eq:rephi}) we obtain the following relation for
the rescaled functions $\phi$
\begin{equation}
  \label{eq:crephi}
 \phi^2_M+  \phi^2_S-   \phi^2_K=-\frac{1}{4\Omega}.  
\end{equation}
Then, the function $\phi_K$ has the form 
\begin{equation}
  \label{eq:phiK2}
   \phi_K=\frac{f_K}{\sqrt{\Omega}}, 
   \quad f_K=\sqrt{\Omega(\phi_M^2+\phi_S^2)+1/4}.
\end{equation}
The  function $f_K$ is analytic. We see that $\phi_K$ is not
analytic, in fact it blows up at $i$.  The 
functions $\omega$ and $\lambda$  are
given by
\begin{equation}
  \label{eq:clt}
  \omega =\frac{\phi_S}{(\phi_K-\phi_M)}, \quad \lambda =
  \frac{1}{2\sqrt{\Omega} (\phi_K-\phi_M)}.
\end{equation}
From these formula, we see immediately that these functions are also not
analytic at $i$, since $\sqrt{\Omega}$ is not analytic. 

From  lemma  \ref{Eo} and equations (\ref{eq:clt}), (\ref{eq:phiK2})  it follows that
\begin{equation}
  \label{eq:El}
  \lambda  \in E^\omega, \quad \lambda(i)=1,
\end{equation}
and 
\begin{equation}
  \label{eq:Eo}
   \omega=r\hat \omega, \quad  \hat\omega \in E^\omega, \quad
   \hat\omega=S^ix_i+O(r^2).
\end{equation}

In order to characterize the behavior of $\beta_i$ we have to solve
equation (\ref{eq:twist2}). First we will write this equation in terms 
of the functions  $\phi$ . We write equation (\ref{eq:fieldphi}) as 
\begin{equation}
  \label{eq:fieldphi1}
  L_{\tilde \gamma} \tilde \phi= \frac{15}{8}\tilde \phi,
\end{equation}
where $L_{\tilde \gamma}=\tilde \Delta - \tilde R/8$.  We use the
formula 
\begin{equation}
  \label{eq:relL}
L_{\tilde \gamma} \tilde \phi  = \Omega^{5/2} L_{\gamma}\phi, 
\end{equation}
which holds for an arbitrary conformal factor, to obtain
\begin{equation}
  \label{eq:fieldphi2}
  L_{\gamma}  \phi= \frac{15}{8}\Omega^{-2} \tilde R  \phi.
\end{equation}
Take equation (\ref{eq:fieldphi2}) for $\phi_M$ and for $\phi_S$, multiply it by
$\phi_S$ and  $\phi_M$ respectively, and take the difference between
this two equations. We obtain
\begin{equation}
  \label{eq:MS}
  \phi_S \Delta \phi_M-  \phi_M \Delta \phi_S=0.
\end{equation}
From this equation it follows that the vector $J_i^1$ defined by 
\begin{equation}
  \label{eq:J1}
 J_i^1=\phi_S D_i \phi_M - \phi_M D_i \phi_S,
\end{equation}
is divergence free $ D^iJ_i^1=0$. In the same way we obtain that the
following vectors are divergence free 
\begin{align}
  \label{eq:J2}
  J_i^2 &=\phi_M D_i \phi_K - \phi_K D_i \phi_M,\\
 \label{eq:J3}
  J_i^3 &=\phi_S D_i \phi_K - \phi_K D_i \phi_S.
\end{align}
The vector $J^1$ is analytic, but the vectors $J^2$ and $J^3$ are
not. In particular, $J^3$ has the form 
\begin{equation}
  \label{eq:aj3}
    J_i^3=\frac{1}{\Omega^{3/2}}\left \{\Omega^2( \phi_S D_i f_K 
  -f_KD_i \phi_S) - \frac{1}{2}\phi_S f_K D_i \Omega \right \}.
\end{equation}
The expression  in curly brackets in this equation is analytic.

In terms of the conformal metric $\gamma_{ij}$ equation (\ref{eq:twist2})
is given by 
\begin{equation}
  \label{eq:twistc}
\frac{1}{\Omega \lambda^2} D_i \omega= 4(J_i^1 -J_i^3) = -  \epsilon_{ijk}  D^j
  \beta^k , 
\end{equation}
where $\epsilon_{ijk}$ is the volume element with respect to
$\gamma_{ij}$, and we have used (\ref{eq:clt}). This equation can be
written in terms of the flat metric $\delta_{ij}$, the flat volume
element $\bar \epsilon_{ijk}$ and partial derivatives $\partial_i$
\begin{equation}
  \label{eq:twistf}
 \bar \epsilon^{ijk}  \partial_j
  \beta_k=J^i, \quad J^i= - \frac{4}{\sqrt{|\gamma|}}
  \gamma^{ij} (J_j^1 -J_j^3). 
\end{equation}
Note that $\partial_i J^i=0$. We use
equation (\ref{eq:aj3}) and lemma \ref{r2} to
conclude that the vector $J^i$, in normal coordinates, has the form
\begin{equation}
  \label{eq:ji}
  J^i=H_1^i+\frac{1}{r^3} (r^2 H_2^i+ x^i H),
\end{equation}
where $H_1^i, H_2^i, H$ are analytic functions. Using (\ref{eq:sme})
we find that
\begin{equation}
  \label{eq:H2e}
 H_2^i(0)=S^i.  
\end{equation}
 
\begin{lemma} \label{beta}
There exist a solution $\beta_i$ of equation (\ref{eq:twistc}) which, 
  in normal coordinates $x^i$, has the following form
  \begin{equation}
    \label{eq:lbeta}
    \beta_i= \beta^1_i+\frac{\beta^2_i}{r},
  \end{equation}
where $\beta^1_i$ and   $\beta^2_i$ are analytic functions of
$x^i$ given by 
\begin{equation}
    \label{eq:lbeta12}
\beta^1_i=\bar\epsilon_{ijk} f_1^j x^k, \quad
\beta^2_i=\bar\epsilon_{ijk} f_2^j x^k, 
  \end{equation}
where $f_1^i$ and  $f_2^i$ are analytic. In particular, this implies
that $\beta_i x^i=0$.
\end{lemma}
\begin{proof}
  We expand in powers series in the coordinates $x^i$ the analytic functions $H_1^i, H_2^i, H$
  in the expression (\ref{eq:ji}). For each power, we use the
  following explicit formula in order to solve (\ref{eq:twistf}).  Let
  $p^i_{(m)}$ a three tuple of homogenous polynomials of order $m$,
  which satisfies 
\begin{equation}
  \label{eq:rs}
  \partial_i (r^s p^i_{(m)})=0,
\end{equation}
for some integer  $s$. Then, we have
\begin{equation}
  \label{eq:bp}
  \bar \epsilon^{ijk} \partial_j \beta^{(m)}_k=r^s p^i_{(m)}, \quad
  \beta^{(m)}_i =r^s (s+m+2)^{-1}\bar \epsilon_{ijk} p_{(m)}^j x^k. 
\end{equation}
By (\ref{eq:bp}), the series defined by the $p^i_{(m)}$ majorizes the
one defined by the $\beta^{(m)}_k$, hence the last one defines a
convergent power series.  Note that the term with $x^i$ in (\ref{eq:ji})
 made no contribution to $\beta_i$.
\end{proof}

We define the conformal factor 
\begin{equation}
  \label{eq:confs}
  \hat \Omega=\sqrt{\lambda} \Omega,
\end{equation}
and the rescaled four  metric and three metric by  $g_{ab}=
\hat{\Omega}^2\tilde g_{ab}$ and
$h_{ij} = \hat \Omega^2 \tilde h_{ij}$. Note that $\hat \Omega $ is
not analytic. 
We have
\begin{equation}
  \label{eq:cstmetric}
   g = \Omega^2\lambda^2(dt+\beta_i d  x^i)^2- \gamma_{ij}d  x^i
  d  x^j.
\end{equation}
The metric $g_{ab}$ has also a $3+1$ decomposition with respect to the
hypersurface $t=constant$
\begin{equation}
  \label{eq:3+1c}
  g =  N^2 dt^2-  h_{ij} (N^i dt+  x^i )(N^j
  dt+ d  x^j). 
\end{equation}
We define
\begin{equation}
  \label{eq:sbc}
  N_j=  h_{ij}N^i, \quad  \beta^j=  \gamma^{jk} \beta_k.
\end{equation}
Then we have the followings relations
\begin{equation}
  \label{eq:N}
  N=\hat {\Omega} \tilde N, \quad  N=\frac{\Omega^2
    \lambda^2}{(1-\Omega^2\lambda^2\beta_j\beta^j)}, \quad  \beta^j=-\frac{N^j}{N^2}, 
\end{equation}
and 
\begin{equation}
  \label{eq:chh}
  h_{ij}=\gamma_{ij}-\Omega^2\lambda^2 \beta_i \beta_j.
\end{equation}
Note that
\begin{equation}
  \label{eq:Ncc}
  N=r^2 f_N,\quad f_N \in E^\omega,\quad f_N(0)=1.
\end{equation}
From equation (\ref{eq:chh}),   (\ref{eq:El}), (\ref{eq:Eo}), lemma
\ref{r2}, lemma \ref{beta} and theorem \ref{bs}, we can read off the  following
theorem.
\begin{theorem}
 \label{tm}
 Assume $\beta$  given by lemma \ref{tm}. Then, in some neighborhood of
 $i$, the metric $h_{ij}$ has the form
  \begin{equation}
    \label{eq:th3}
  h_{ij}=  h^1_{ij}+r^3 h^2_{ij},  
  \end{equation}
  where $h^1_{ij}$ and $h^2_{ij}$ are analytic.
\end{theorem}

We remark that $ h_{ij} \in W^{4,p}$, $p<3$ (see e.g. \cite{Adams} for
the definitions of the Sobolev and H\"older spaces $W^{s,p}$ and
$C^{m,\alpha}$). This follows from expression (\ref{eq:th3}) and $r^3
\in W^{4,p}$, $p<3$. It implies, in particular, that the metric is in
$C^{2,\alpha}$.  With the conformal factor (\ref{eq:confs}), we can
define the conformal compactification $S$ of the Cauchy slice $\tilde
S$ plus the point at infinity $i$, in the same way as we made for
$\tilde X$.  Theorem \ref{tm} says that $\tilde S, \tilde h$ admit a
$C^{2,\alpha}$ compactification. Of course, theorem \ref{tm} only
describe the behavior of the fields in a particular  foliation.  
However, happens very  unlikely that for other foliations the smoothness
improve.
The decomposition \eqref{eq:lbeta} for $\beta^k$, and hence the
decomposition \eqref{eq:th3} for $h_{ij}$, is preserved under the
transformation
\begin{equation}
  \label{eq:2}
  \beta_k\rightarrow \beta_k +\partial_k f, \quad f\in E^\omega.
\end{equation}
If we impose the condition $x^k\beta_k=0$, then $\partial_k f$ is
fixed. That is, $\beta^k$ given by lemma \ref{beta} is the unique vector that
satisfies both equation \eqref{eq:twistf} and  $x^k\beta_k=0$. 

Let  $\tilde \chi_{ij}$ be the extrinsic curvature of the
hypersurface $\tilde S$ with respect to the metric $\tilde g_{ab}$. We
denote by  $\chi_{ij}$ the
 extrinsic curvature of the same hypersurface, but with respect to the
 conformal metric $g_{ab}$. These two tensors  are related by $\tilde
 \chi_{ij}= \Omega^{-1} \chi_{ij}$. This formula is valid since the
 conformal factor $\Omega$ is independent of the coordinate $t$.  
We define the tensor $\psi_{ij}$
 by
\begin{equation}
  \label{eq:tcp}
\psi_{ij}=\hat\Omega^{-1}\tilde \chi_{ij}= \hat\Omega^{-2} \chi_{ij}.
\end{equation}
The tensor $\psi_{ij}$ plays  an important role in the conformal method for
solving the constraint equations. The following theorem characterizes
the behavior of $\psi_{ij}$ near $i$.

\begin{theorem}
 \label{tc}
Assume $\beta$   given by lemma \ref{tm}. Then, in some neighborhood of $i$, the tensor $\psi_{ij}$ has the following form
  \begin{equation}
    \label{eq:chit}
    \psi_{ij}=r^{-5}f  x_{(i}\beta_{j)}^2 +r^{-3}\hat
    \psi_{ij},  
  \end{equation}
  where $\hat \psi_{ij}, f \in E^\omega$ and the analytic vector
  $\beta_j^2$ is given by (\ref{eq:lbeta12}). Moreover,
  $r^8\psi_{ij}\psi^{ij}\in E^\omega$.
\end{theorem}
\begin{proof} Since $h_{ij}$ does  not depend on $t$, the extrinsic
curvature $\chi_{ij}$ satisfies the equation
\begin{equation}
  \label{eq:cchi}
  \chi_{ij}=-\frac{1}{2N}\pounds_{N^k} h_{ij},
\end{equation}
where $\pounds_{N^k}$ denote the Lie derivative with respect to the
vector field $N^k$. We express this equation in terms of 
 $D_i$, the covariant derivative
with respect to $\gamma_{ij}$ 
\begin{align}
  \label{eq:chic2}
 \chi_{ij} &=-\frac{1}{2N} (N^k D_k h_{ij} +2h_{k(i}D_{j)}N^k) \\
     &=-\frac{1}{2} N\beta^k D_k (\lambda^2 \Omega^2 \beta_i
     \beta_j)+\frac{1}{N}h_{k(i}D_{j)}(\beta^kN^2),   
\end{align}
where, in the second line, we have used equations (\ref{eq:N}) and
(\ref{eq:chh}). Then we use lemma \ref{beta}, equation (\ref{eq:Ncc}),
and equation (\ref{eq:tcp}) to prove (\ref{eq:chit}). To compute
$r^8\psi_{ij}\psi^{ij}$ we use equation (\ref{eq:chit}) and
$x^i\beta^2_i=0$.
\end{proof}

Using equation (\ref{eq:H2e}) we obtain that $\psi_{ij}$ has a
expansion of the form
\begin{equation}
  \label{eq:psiexp}
   \psi_{ij}=r^{-5} 3x_{(i}\bar \epsilon_{j)qk}S^qx^k  +O(r^{-2}),
\end{equation}
where $S^i$ is the angular momentum defined by (\ref{eq:sme}). 

In general these slices will not be maximal, i.e., $ \chi=
h^{ij}  \chi_{ij} \neq 0$. However, if the space-time is axially
symmetric, this foliation can be chosen to be also maximal. In order
to prove this, assume that we have an axial  Killing vector
$\eta^a$. The projection $\eta^i$ of $\eta^a$ into  the hypersurface
$\tilde S$ is a Killing vector of the metric $h_{ij}$. We can
chose a coordinate system adapted to $\eta^a$. In these coordinates we
have (see for example \cite{Beig00b}) $N_i=\sigma \eta_i$, where
$\pounds_\eta \sigma=0$. Then, using equation (\ref{eq:cchi}) and the
Killing equation, we obtain
\begin{equation}
  \label{eq:max}
 \chi_{ij}=-\frac{1}{2 N}\eta_{(i}D_{j)}\sigma.
\end{equation}
From this equation we deduce that $\chi=0$. The same argument applies, 
 of
course,  to  the physical extrinsic curvature $\tilde \chi_{ij}$.

As  an  application of  our results  we fill a gap in the proof made
in \cite{Damour90} of the following theorem.
\begin{theorem}[Damour-Schmidt]
 Every stationary, asymptotically flat, vacuum
space-time admits an analytic conformal extention through null infinity. 
\end{theorem}
For the definition of null infinity see \cite{Penrose86}, see also the
review \cite{Frauendiener00}. 

\begin{proof}
  By lemma \ref{beta} and equation (\ref{eq:clt}) we have that
  $\beta_i$ and $\lambda$ are analytic with respect to $r$ and the
  angles $x^i/r$ (but not with respect to $x^i$!).  This proves the
  assumption made in \cite{Damour90}, $\lambda$ and $\beta_i$ are
  essentially the functions $F$ and $F_\alpha$ defined there.
\end{proof}

\section{Final Comments}

The vector $\beta^k$ is the essential piece in the translation from
the quotient manifold $\tilde X$ to the Cauchy slice $\tilde S$. This
vector is computed as follows.  We calculate first the vector $J^i$ by
equation \eqref{eq:twistf} in terms of  the analytic
potentials $\phi_S$, $\phi_M$ and the analytic metric $\gamma_{ij}$.
The vector $J^i$ is divergence free with respect to the flat
metric. This vector is not analytic in terms of the Cartesian
coordinates $x^i$, since the radial function $r$ appears explicitly in
it. The curl of $\beta^k$ is $J^i$ (cf. equation
\eqref{eq:twistf}). In lemma \ref{beta} we found a solution of this
equation which has the desired properties: although is not analytic in
$x^i$ it has an analytic expansion in terms of $r$ and the
corresponding angles $x^i/r$. Given the multipole expansion of
$\phi_S$, $\phi_M$ and  $\gamma_{ij}$ one can compute the
corresponding expansion for $\beta^k$ with equation
\eqref{eq:bp}. Using  lemma \ref{beta} it is straightforward to prove
our main result given by theorems \ref{tm} and \ref{tc}. Those theorems
 characterize the behavior of the
metric and the extrinsic curvature of $\tilde S$ near infinity. In
particular, we prove that the intrinsic metric $h_{ij}$ is not
analytic at infinity, unless we are in the static case. 

The Kerr metric is a particular
important example of a stationary space time. Explicit computation for
this metric has been made in \cite{Dain00c}, in which we see the
behavior described by theorems \ref{tm} and \ref{tc}.

The condition $r^8\psi_{ij}\psi^{ij} \in E^\omega$ has been used in
\cite{Dain99} to construct general initial data which have asymptotic
expansion in terms of powers of the radial coordinate. In this work,
it has been assumed that the conformal metric is smooth
at $i$. Theorem \ref{tm} suggest that the same result holds if we only
require that the metric has the form (\ref{eq:th3}).  This
generalization will be studied in a subsequent  work \cite{Dain01}.

\section*{Acknowledgements} 
I would like to thank R. Beig and M. Mars for many interesting
discussions and suggestions.  Especially,  I would like to thank
H. Friedrich for sharing with me his private calculations and for a
carefully reading of the original manuscript.

%\bibliographystyle{abbrv}
%\bibliography{biblio}

\begin{thebibliography}{10}

\bibitem{Adams}
R.~A. Adams.
\newblock {\em Sobolev Spaces}.
\newblock Academic Press, New York, 1975.

\bibitem{Beig91b}
R.~Beig.
\newblock Conformal properties of static spacetimes.
\newblock {\em Classical and Quantum Gravity}, 8(2):263--271, 1991.

\bibitem{Beig00b}
R.~Beig and B.~Schmidt.
\newblock Time-independent gravitational fields.
\newblock In B.~Schmidt, editor, {\em {E}instein's Field Equations and Their
  Physical Implications}, volume 540, pages 325--372. Springer Lecture Note in
  Physics, 2000.
\newblock Also in gr-qc/0005047.

\bibitem{Beig81}
R.~Beig and W.~Simon.
\newblock On the multipole expansion for stationary space-times.
\newblock {\em Proc. R. Lond. A}, 376:333--341, 1981.

\bibitem{Dain00c}
S.~Dain.
\newblock Initial data for two {K}err-like black holes.
\newblock Submitted for publication, gr-qc/0012023, 2000.

\bibitem{Dain01}
S.~Dain.
\newblock Asymptotically flat initial data with prescribed regularity {II}.
\newblock In preparation., 2001.

\bibitem{Dain99}
S.~Dain and H.~Friedrich.
\newblock Asymptotically flat initial data with prescribed regularity.
\newblock Accepted for publication in Comm. Math. Phys. , gr-qc/0102047, 2000.

\bibitem{Damour90}
T.~Damour and B.~Schmidt.
\newblock Reliability of perturbation theory in general relativity.
\newblock {\em J. Math. Phyys.}, 31(10):2441--2453, 1990.

\bibitem{Frauendiener00}
J.~Frauendiener.
\newblock Conformal infinity.
\newblock {\em Living Reviews in Relativity}, 2000.

\bibitem{Friedrich88}
H.~Friedrich.
\newblock On static and radiative space-time.
\newblock {\em Commun. Math. Phys.}, 119:51--73, 1988.

\bibitem{Geroch71}
R.~Geroch.
\newblock A method for generating solutions of {E}instein's equations.
\newblock {\em J. Math. Phys.}, 12(6):918--924, 1971.

\bibitem{Hansen74}
R.~O. Hansen.
\newblock Multipole moments of stationary space-times.
\newblock {\em J. Math. Phys.}, 15(1):46--52, 1974.

\bibitem{Kennefick95}
D.~Kennefick and N.~O. Murchadha.
\newblock Weakly decaying asymptotically flat static and stationary solutions
  to the einstein equations.
\newblock {\em Classical and Quantum Gravity}, 12(1):149--158, 1995.

\bibitem{Kundu81}
P.~Kundu.
\newblock On the analyticity of stationary gravitational fields at spatial
  infinity.
\newblock {\em J. Math. Phys.}, 22(9):2006--2011, 1981.

\bibitem{Penrose86}
R.~Penrose and W.~Rindler.
\newblock {\em Spinors and Space-Time}, volume~2.
\newblock Cambridge University Press, Cambridge, 1986.

\bibitem{Reula89}
O.~Reula.
\newblock On existence and behaviour of asymptotically flat solutions to the
  stationary {E}instein equations.
\newblock {\em Commun. Math. Phys.}, 122:615--624, 1989.

\end{thebibliography}

\end{document}